\documentclass[12pt]{article}
\usepackage[a4paper, left=1.2in, right=1.2in, top=1in, bottom=1in]{geometry}
\usepackage{caption}

\usepackage{url} 
\usepackage{lineno}
\usepackage{soul}
\usepackage{float}
\usepackage{apacite} 
\usepackage{authblk}
\usepackage{natbib}
\usepackage{graphicx}
\usepackage{xcolor}
\usepackage[utf8]{inputenc}
\usepackage{orcidlink}
\usepackage{titlesec}

\titleformat{\section}
  {\large\bfseries}  
  {\thesection}  
  {1em}  
  {}

\titleformat{\subsection}
  {\normalsize\bfseries}  
  {\thesubsection}  
  {1em}  
  {}

\title{\textbf{Quasi-Periodic Pulsations in Ionospheric TEC Synchronized with Solar Flare EUV Emission}}

\author{%
    \textbf{Aisling N. O'Hare\orcidlink{0009-0004-8025-9545}}\textsuperscript{1}, 
    \textbf{Susanna Bekker}\textsuperscript{1}, 
    \textbf{Laura A. Hayes}\textsuperscript{2,3}, 
    \textbf{Ryan O. Milligan}\textsuperscript{1}
}
\date{}

\begin{document}

\maketitle
\vspace{-0.1\textwidth}
\small{\begin{center}
    \textsuperscript{1}Astrophysics Research Centre, School of Mathematics and Physics, Queen’s University Belfast, University Road, Belfast BT7 1NN, UK\\
    \textsuperscript{2}European Space Agency, ESTEC, Keplerlaan 1-2201 AZ, Noordwijk, The Netherlands\\
    \textsuperscript{3}Astronomy \& Astrophysics Section, School of Cosmic Physics, DIAS Dunsink Observatory, Dublin D15XR2R, Ireland
\end{center}}


\section*{\normalsize Key Points}
\begin{itemize}
    \item Synchronous pulsations in EUV solar flare emission and ionospheric total electron content (TEC) were detected.
    \item Wavelet and periodogram analyses were applied to EUV flux data and TEC from 251 GPS stations during an X5.4 solar flare.
    \item The period of synchronous solar and ionospheric variations was approximately 85\,s, with an average time delay of 30\,s.
\end{itemize}
\newpage
\section*{Abstract}
The extreme ultraviolet (EUV) and X-ray radiation emitted during solar flares has been shown to significantly increase the electron density of the Earth's ionosphere. During flares, quasi-periodic pulsations (QPPs) in X-ray flux originating in the corona have previously been linked to subsequent pulsations in the Earth's ionospheric D-region. Similar pulsations have been detected in chromospheric EUV emission, although their impact on the Earth's ionosphere has not previously been investigated. Here, for the first time, synchronous pulsations were detected in solar EUV emission and ionospheric Total Electron Content (TEC) measurements. Using wavelet and periodogram analysis, we detect QPPs with approximately 85 second periods in chromospheric EUV emission lines (He~{\sc{ii}} 304\,\AA{}, C~{\sc{iii}} 977\,\AA{} and H~{\sc{i}} 972\,\AA{}) from the Solar Dynamics Observatory Extreme Ultraviolet Variability Experiment (SDO/EVE) during the impulsive phase of an X5.4 flare on March 7, 2012. These lines contribute to ionization in the ionospheric E- and F-regions, resulting in subsequent variations of electron density with the same periodicity, which was detected in TEC measurements. This work demonstrates that the Earth's ionosphere is responsive to fine-scale fluctuations in EUV emission during flares, with a time delay of approximately 30\,seconds found. These findings may have applications in atmospheric modeling and solar-terrestrial studies, including the calculation of ionospheric recombination rates.

\section*{Plain Language Summary}
Solar flares release vast amounts of energy, primarily as extreme ultraviolet (EUV) and X-ray emission, which can significantly increase electron density in Earth’s ionosphere. In the past, periodic fluctuations in X-ray radiation from the Sun’s corona were linked to similar variations in the Earth’s lower ionosphere. This study investigates whether pulsations in chromospheric EUV emissions also affect Earth's ionosphere, using data from a powerful X5.4 flare on March 7, 2012. For the first time, synchronized pulsations with periods of approximately 85 seconds were observed in EUV emissions and ionospheric Total Electron Content (TEC), with a time delay of about 30 seconds. This finding suggests that even small fluctuations in EUV radiation during flares can impact Earth’s ionosphere, with potential applications in atmospheric modeling and studying interactions between the Sun and Earth. 

\section{Introduction}

During solar flares, significant amounts of energy are released resulting in radiation emitted from the Sun across the entire electromagnetic spectrum. A common feature of flare emission observed in various wavelengths is quasi-periodic pulsations (QPPs). QPPs are generally described as regular pulsations or fluctuations in solar flare emission that have a periodic or characteristic timescale. They are a widely observed feature of solar flare emission, with some studies suggesting that at least 20\% of solar flares exhibit evidence of stationary QPPs \citep{pugh2017properties}, while \citet{simoes2015soft} reported that 80\% of X-class flares during Solar Cycle 24 displayed QPPs in soft X-ray (SXR) emission during their impulsive phases. In comparison, \citet{dominique2018detection} concluded that 81/90 flares of class $>$M5.0 studied showed QPPs in EUV emission, and in a study of 675 M- and X-class flares, \citet{inglis2016large} found that approximately 30\% showed strong signatures of QPPs in SXR emission. Additionally, \citet{hayes2020} found that approximately 46\% of X-class, 29\% of M-class, and 7\% of C-class flares in Solar Cycle 24 show evidence of stationary QPPs in SXRs. It should be noted, that while these studies show different rates of QPPs, this depends on whether the method used was looking for stationary or non-stationary QPPs. Stationary QPPs are those that show a strong periodic component, with consistent period and phase over time, while non-stationary QPPs show variable time spacing between oscillations, which may increase or decrease systematically \citep{kupriyanova2010types}. Either way, the presence of pulsations in solar flares, are extremely common. QPPs were first observed in solar flares over 55 years ago, when \citet{parks1969sixteen} reported a 16\,s periodic fluctuation in hard X-ray (HXR) flare emission. Since then, studies of QPPs in HXR flare emission have reported periods ranging from sub-second to several minutes \citep{Nakariakov2009, dennis2017detection, knuth2020subsecond, hayes2020, collier2024localising, Inglis2024}. Soft X-ray, ultraviolet (UV) and extreme ultraviolet (EUV) emissions during solar flares have been found to exhibit QPPs with periods ranging from tens of seconds up to $\simeq$5\,min \citep{dolla2012time, brosius2015quasi, simoes2015soft, li2015imaging, brosius2016quasi, tian2016global, ning2017one, Milligan2017}. Additionally, longer periods, extending up to approximately 20 minutes, have been reported in SXR and EUV \citep{Hayes2017}.

The mechanisms responsible for driving QPPs in solar flares remain a focus of intensive theoretical studies. For comprehensive reviews of such models, see \citet{Zimovets2021}, \citet{mclaughlin2018modelling} and \citet{VanDoorsselaere2016}. These models explore various mechanisms, including the modulation of plasma by MHD (magnetohydrodynamic) oscillations in flaring loops, periodic energy release driven by MHD modes, and intrinsically periodic energy release processes. Some observational studies suggest that this periodic behavior may result from the quasi-periodic injection of non-thermal electrons down into the chromosphere by episodic magnetic reconnection \citep{brosius2016quasi, collier2024localising}. While QPPs in solar flare emission have been observed for many years, their potential to induce similar behavior in the Earth's ionosphere has not been fully investigated and it remains unclear how sensitive the ionosphere is to small scale changes in solar flare EUV emission. As QPPs are an inherent characteristic of solar flares, understanding their geophysical impact is crucial.

Flare emission, that occurs on the Earth-facing side of the Sun, can cause an increase in electron density in different layers of the Earth's ionosphere \citep{mitra1974ionospheric}. In particular, X-ray ($<$100\,\AA{}) and EUV (100–1000\,\AA{}) emission from solar flares affect the entire dayside ionosphere \citep{tsurutani2009brief}, dramatically increasing the ionization and molecular dissociation of atmospheric components at different altitudes \citep{wan2005gps}. Generally, EUV flare emission disturbs the middle (E-region; 90–150\,km) and upper ionosphere (F-region; 150–200\,km) where the electron density is high. In contrast, higher energy soft and hard X-ray photons can penetrate to the lower lying D-region of the ionosphere (60–90\,km). During quiet Sun conditions, the D-region is formed and maintained through the ionization of nitric oxide (NO) by solar Lyman-$\alpha$ emission (1216\,\AA{}). However, during solar flares, X-ray emission ($<$10\,\AA{}) increases by many orders of magnitude, causing an increase in ionization of the main neutral components, O$_2$ and N$_2$ \citep{mitra1974ionospheric}. These molecules are of much higher abundances than NO and therefore their ionization is the main driver of the observed increase in electron density in the D-region during flares. These changes lead to perturbations of very low frequency (VLF) signals (3–30 kHz), which propagate in the waveguide between the Earth’s surface and the ionospheric D region \citep{thomson2001solar, raulin2013response, hayes2021solar, nina2021modelling, bekker2023influence} and can cause radio blackouts. In the E-layer, softer X-rays (10–100\,\AA{}) are primarily absorbed by O$_2$ and N$_2$, while in the F-layer, EUV photons are the dominant source of ionization of atomic oxygen (O). Therefore, the sudden increase in solar irradiance caused by solar flares induces compositional changes across the entire dayside ionosphere of the Earth, which can significantly influence communication and navigation systems. 

Total Electron Content (TEC) is defined as the total number of electrons integrated between two points along a tube of 1m$^2$ cross section. The remote sensing method for calculating TEC derived from the Global Positioning System (GPS) has become a significant tool for monitoring fluctuations in ionospheric electron density. The majority of the electron content in the ionosphere is located in the most ionized F-region, which is why EUV flare emission predominantly causes changes in TEC \citep{leonovich2002estimating, tsurutani2009brief}. These perturbations affect the code and phase delays of the received GPS signals, which can be used to monitor the state and dynamics of the ionosphere, and quantify changes in TEC during solar flares \citep{wan2002sudden, wan2005gps, garcia2007solar, yasyukevich20186}.

Previously, \citet{Hayes2017} reported that a series of X-ray pulsations of GOES class B9.2–C6.8 induced synchronous pulsations in VLF measurements of the ionospheric D-region, with a time delay of approximately 90s. This leads to the question of whether pulsations in other wavelengths of flare emission (e.g. EUV) will have a similar effect on other layers of the Earth's ionosphere. In this paper, we present the observations of an X-class flare on March 7, 2012, that exhibited quasi-periodic behavior in EUV emission during its impulsive phase. The purpose of this study is to investigate synchronous pulsations in EUV solar flux and TEC measurements using wavelet analysis and periodogram significance testing to identify QPPs. In Section 2, the solar and ionospheric data used for this study are described. Section 3 outlines the methods of periodicity detection employed; wavelet analysis (Section 3.1) and periodogram significance testing (Section 3.2), as well as cross-correlation analysis (Section \ref{crosscorr}). Section 4 discusses the key results.

\section{Observations and Data Selection}
\subsection{Solar Observations}\label{solobs}
\begin{figure}  
    \centering
    \includegraphics[width=.8\linewidth]{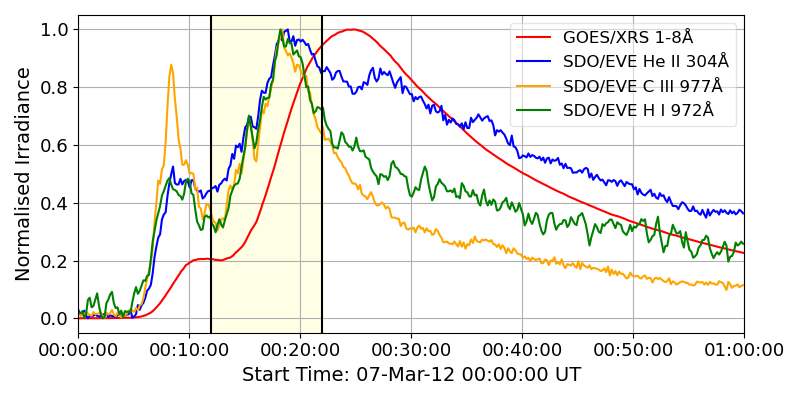}
    \caption{Flare lightcurves in the GOES/XRS 1–8\,\AA{} channel (red), and the three EUV emission lines He~{\sc{ii}} 304\,\AA{} (blue), C~{\sc{iii}} 977\,\AA{} (orange) and H~{\sc{i}} 972\,\AA{} (green) as measured by SDO/EVE. The interval during which QPPs were found (00:12–00:22\,UT) is marked by the yellow shaded area in between the two solid black vertical lines.}
    \label{norm_lightcurves}
\end{figure}

The flare examined in this study was an X5.4 flare that occurred on March 7, 2012 from solar active region NOAA (National Oceanic and Atmospheric Administration) 11429. This flare began at 00:02\,UT and peaked in soft X-rays at 00:24\,UT. EUV spectral irradiance observations of this flare were recorded by the EVE (Extreme ultraviolet Variability Experiment) instrument \cite{woods2012extreme} on board the SDO (Solar Dynamics Observatory) satellite \cite{pesnell2012solar}, with a cadence of 10\,s. Figure \ref{norm_lightcurves} shows the normalized SXR lightcurve measured in the 1–8\,\AA{} channel of the XRS (X-ray Sensor) on board GOES-15 (Geostationary Operational Environmental Satellite), as well as lightcurves from three EUV lines measured by EVE: He~{\sc{ii}} 304\,\AA{}, C~{\sc{iii}} 977\,\AA{} and H~{\sc{i}} 972\,\AA{} (Lyman-$\gamma{}$).  

\begin{figure}
    \centering
    \includegraphics[width=1\textwidth]{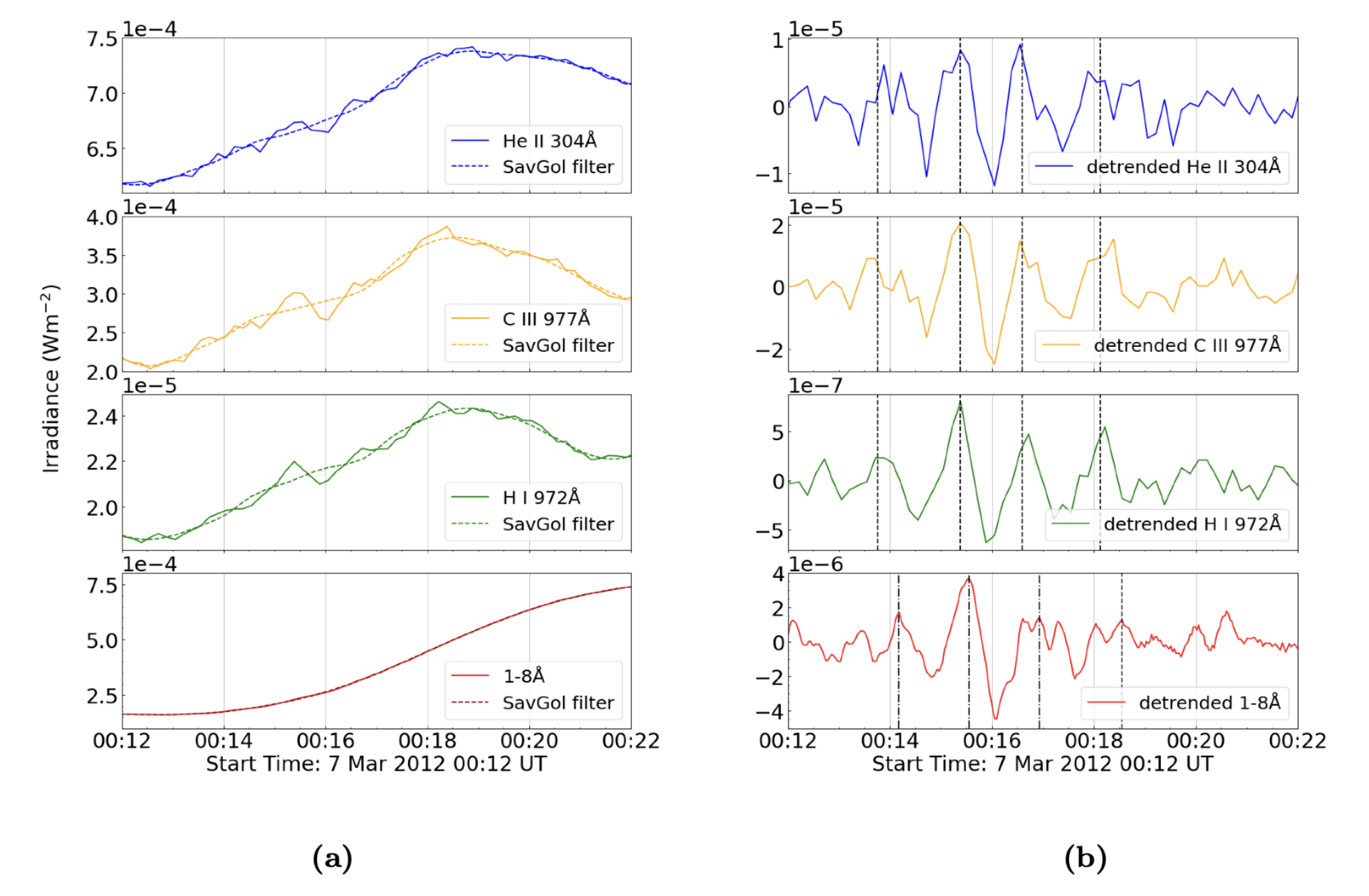}
    \caption{EUV and SXR lightcurves before and after detrending to highlight pulsations. The left panels (\textbf{a}) show the raw solar EUV and SXR lightcurves (top to bottom: He II 304\,\AA{}, C III 977\,\AA{}, H I 972\,\AA{} and 1–8\,\AA{}). The right panels (\textbf{b}) show the three detrended EUV emission line and SXR time series. The vertical dashed lines denote the average peaks of pulsations in the three detrended EUV emission lines. The vertical dashdot lines denote the approximate peaks of pulsations in the detrended SXR lightcurve.}
\label{EUV_detr}
\end{figure}

When selecting EUV emission lines to analyse in this study, the results of theoretical models of the Earth's ionosphere  \citep{Solomon2005, watanabe2021model} were considered. These models enable the calculation of the altitude profile of ionisation rates in the ionosphere during varying levels of solar activity, and theoretically determine the most geoeffective solar radiation lines. Based on these estimates and variations in the spectrum of the flare in this study, three EUV lines were selected for analysis (He~{\sc{ii}} 304\,\AA{}, C~{\sc{iii}} 977\,\AA{} and H~{\sc{i}} 972\,\AA{}), which are most likely responsible for the corresponding increase in the electron content in the ionospheric E and F regions. The irradiance evolution of these lines during the selected flare is shown in Figure \ref{EUV_detr}. All other lines observed by SDO/EVE were examined for QPPs, with many of them (O~{\sc{vi}} 1031.9\,\AA{}, H~{\sc{i}} 1025.7\,\AA{}, H~{\sc{i}} 949.7\,\AA{}, O~{\sc{ii}} 835.5\,\AA{}, O~{\sc{iv}} 790.2\,\AA{}, Ne~{\sc{viii}} 770.4\,\AA{}, Fe~{\sc{xx}} 721.6\,\AA{}, O~{\sc{ii}} 718.5\,\AA{}, O~{\sc{v}} 629.7\,\AA{}, Mg~{\sc{x}} 624.9\,\AA{}, Mg~{\sc{x}} 609.8\,\AA{}, O~{\sc{iii}} 599.6\,\AA{}, He~{\sc{i}} 584.3\,\AA{}, Fe~{\sc{xx}} 567.9\,\AA{}, O~{\sc{iv}} 554.4\,\AA{}, He~{\sc{i}} 537.0\,\AA{}, O~{\sc{iii}} 525.8\,\AA{}, Ne~{\sc{vii}} 465.2\,\AA{}) exhibiting pulsations on the same timescales as the lines selected. However, since these lines are not geoeffective due to their ionization cross sections and weak fluxes, they were omitted.

QPPs can be difficult to detect robustly due to the fact they can be short-lived and have small amplitudes, as well as their quasi-periodic nature. Therefore, to make the small-scale pulsations clear, detrending processes are often applied to remove the overall slowly varying trend and to highlight the fine-scale variations. Each EUV lightcurve was detrended using a Savitzky-Golay \cite{savitzky1964smoothing} filter with a window size of 190\,s, which was subtracted from each original time series to remove the overall trend of the flare. This window size was chosen because it best fit the shape of the flare lightcurves, and it was ensured that changing the window size did not affect the periods of the pulsations present in the time series. The original EUV lightcurves are shown in panels (a) of Figure \ref{EUV_detr} (left). Panels (b) of Figure \ref{EUV_detr} (right) show the detrended EUV lightcurves, where clear pulsations are evident. The vertical dashed lines in panel (b) denote the average time of the peaks in the three EUV emission lines.

Additionally, pulsations on similar timescales as the EUV lines were observed in GOES 1-8\,\AA{} (bottom panels of Figure \ref{EUV_detr}) after detrending with the same window size as the EUV lines. These pulsations appear approximately 10–15\,seconds later than those in EUV, which is expected since SXR emission typically follows impulsive EUV emission during flares. The pulsations in SXR were not analyzed further in this study because this wavelength of flare emission is not a dominant driver of changes in TEC. SXR flare emission typically drives changes in electron density in the D-region, but VLF measurements of the D-region were not available for this flare, and so the impact of the SXR pulsations on the lower ionosphere could not be assessed.

\subsection{Ionospheric Data}
\begin{figure}
    \centering
    \includegraphics[width=.56\textwidth]{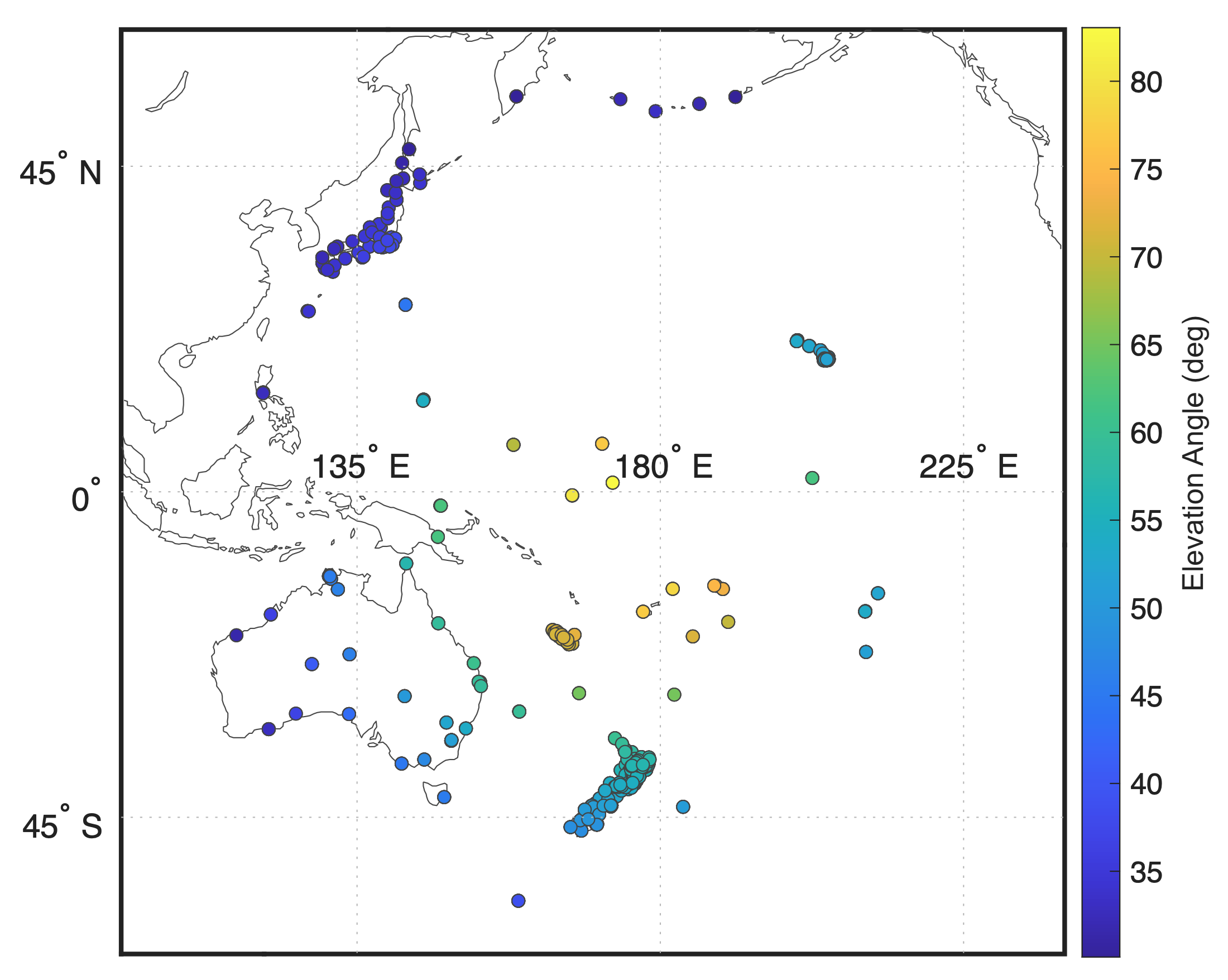}
    \caption{Map showing the location of the GPS stations of the SOPAC network used in this study. The color bar illustrates the solar elevation angle from each station in degrees.}
    \label{map}
\end{figure}

For this study, TEC variations were calculated using GPS data (15\,s cadence) from the Scripps Orbit and Permanent Array Center (SOPAC) network stations. GPS is a system of 24 satellites, divided into six orbital planes across the Earth at a height of $\approx$ 20,200\,km \citep{davies1997studying}, and these satellites use the propagation of very high frequency radio waves to observe ionospheric effects. For further information on how GPS satellites operate, see \citet{HofmannWellenhof1992}. When selecting data for this investigation, we chose stations where the solar elevation angle during the flare was more than 30$^{\circ}$, as enhanced solar emission causes increased ionization in the illuminated part of the Earth's ionosphere. We also limited the latitude range from –55$^{\circ}$ to +55$^{\circ}$ to avoid polar ionospheric phenomena. The locations of the 251 selected stations can be seen in Figure \ref{map}. On average, 4–5 GPS satellites are simultaneously observed by each ground-based station. Using the signal delays on these satellite-receiver beams, we calculated the increase in TEC ($\Delta$TEC) caused by the flare.

\begin{figure}
    \centering
    \includegraphics[width=.66\linewidth]{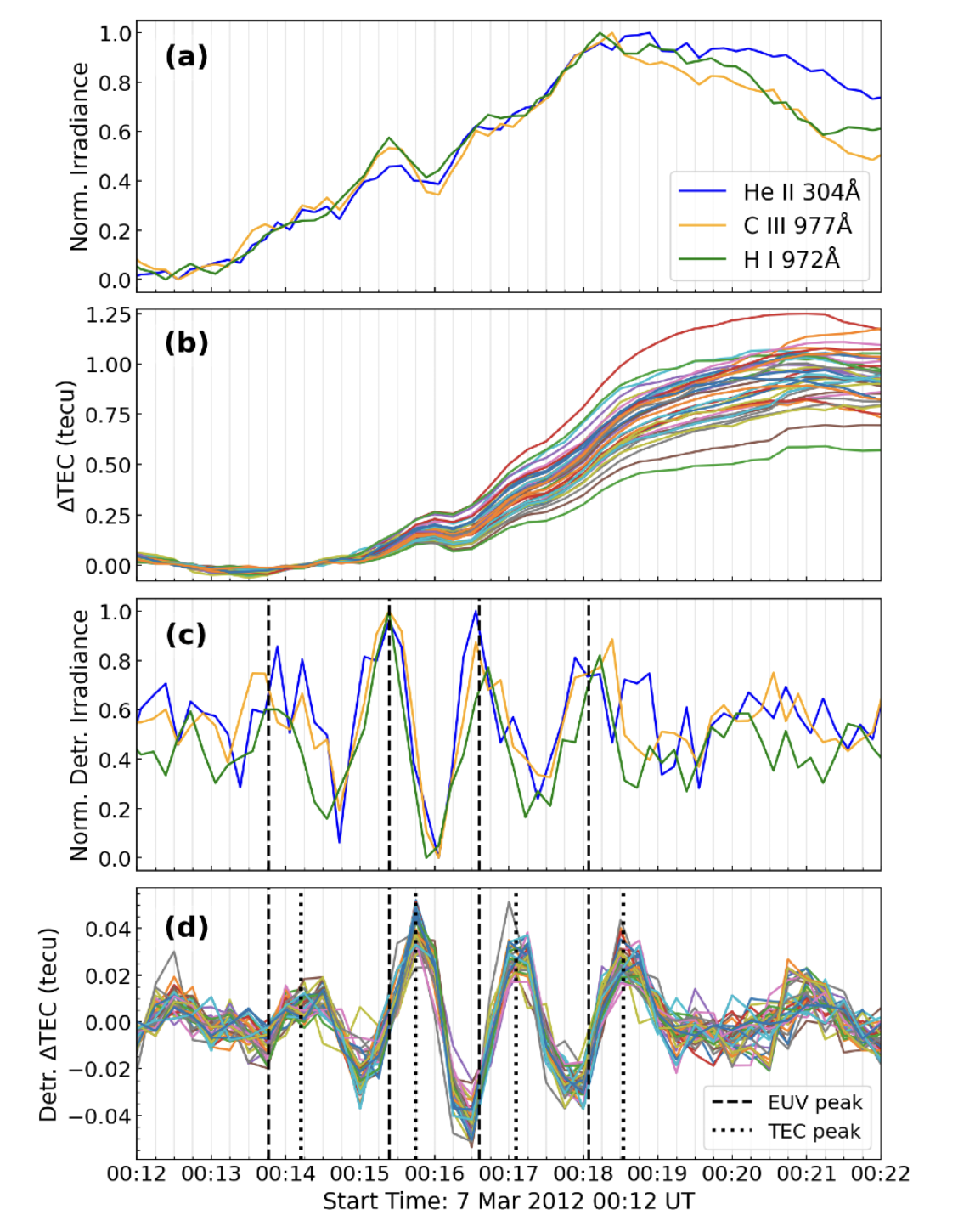}
    \caption{EUV and $\Delta$TEC timeseries before and after detrending to highlight synchronous pulsations. Panel (a) shows the normalized EUV emission lines: He II 304\,\AA{} (blue), C III 977\,\AA{} (orange) and H I 972\,\AA{} (green). Panel (b) shows the $\Delta$TEC. Panels (c) and (d) show the detrended normalized EUV emission lines and the detrended $\Delta$TEC, respectively. The vertical dashed lines denote the average peaks in EUV flare emission, and the vertical dotted lines denote the subsequent peaks in $\Delta$TEC.}
    \label{detrTECEUV}
\end{figure}

Panel (a) of Figure \ref{detrTECEUV} shows the three normalized EUV lightcurves for He II 304\,\AA{}, C III 977\,\AA{} and H I 972\,\AA{}, and the $\Delta$TEC can be seen in panel (b). The normalized, detrended EUV lightcurves are shown in panel (c) of Figure \ref{detrTECEUV}, above the detrended $\Delta$TEC in panel (d). The $\Delta$TEC curves were detrended to remove the overall trend of the flare using the same process as in Section \ref{solobs}; by subtracting a Savitzky-Golay \cite{savitzky1964smoothing} filter with a window size of 190\,s. In Figure \ref{detrTECEUV}, the vertical dashed lines denote the average peaks in EUV, and the vertical dotted lines denote the peaks in $\Delta$TEC. The pulsations in $\Delta$TEC appear approximately 30\,s after those in EUV.

\section{Analysis}

Many periodicity detection methods can yield false detections or fail to detect known pulsations in simulated data \cite{broomhall2019blueprint}. Therefore, it is advantageous to employ more than one detection method to ensure a more robust investigation. QPPs can be identified using a range of methods, including using Fourier transform techniques, direct fitting with a hypothesized oscillatory function, wavelet transforms, as well as many others. Each method employs different ways to estimate false-alarm levels and noise models, as well as having various detection criteria. Based on those presented in statistical studies \cite{broomhall2019blueprint}, for this investigation two methods of periodicity detection were employed: wavelet analysis (Section \ref{wavelet}) and periodogram significance testing (Section \ref{Periodogram}).

\subsection{Wavelet Analysis}
\label{wavelet}
Wavelet analysis is a widely used method of detection for QPPs \citep{dolla2012time, hayes2016quasi, dennis2017detection, dominique2018detection} and is based on the process outlined in \citet{torrence1998practical}. The wavelet power spectrum shows the amount of power that is present at a certain scale (or period) and is used to determine dominant periods that are present in time series. The significance of enhanced power in the wavelet spectra is then tested using a background spectrum. In this study we have assumed a white noise background. A detected period for this study was defined as having a peak in the global power spectrum that lies above the 99$\%$ significance level.

\begin{figure}[H]
    \centering
    \includegraphics[width=1\linewidth]{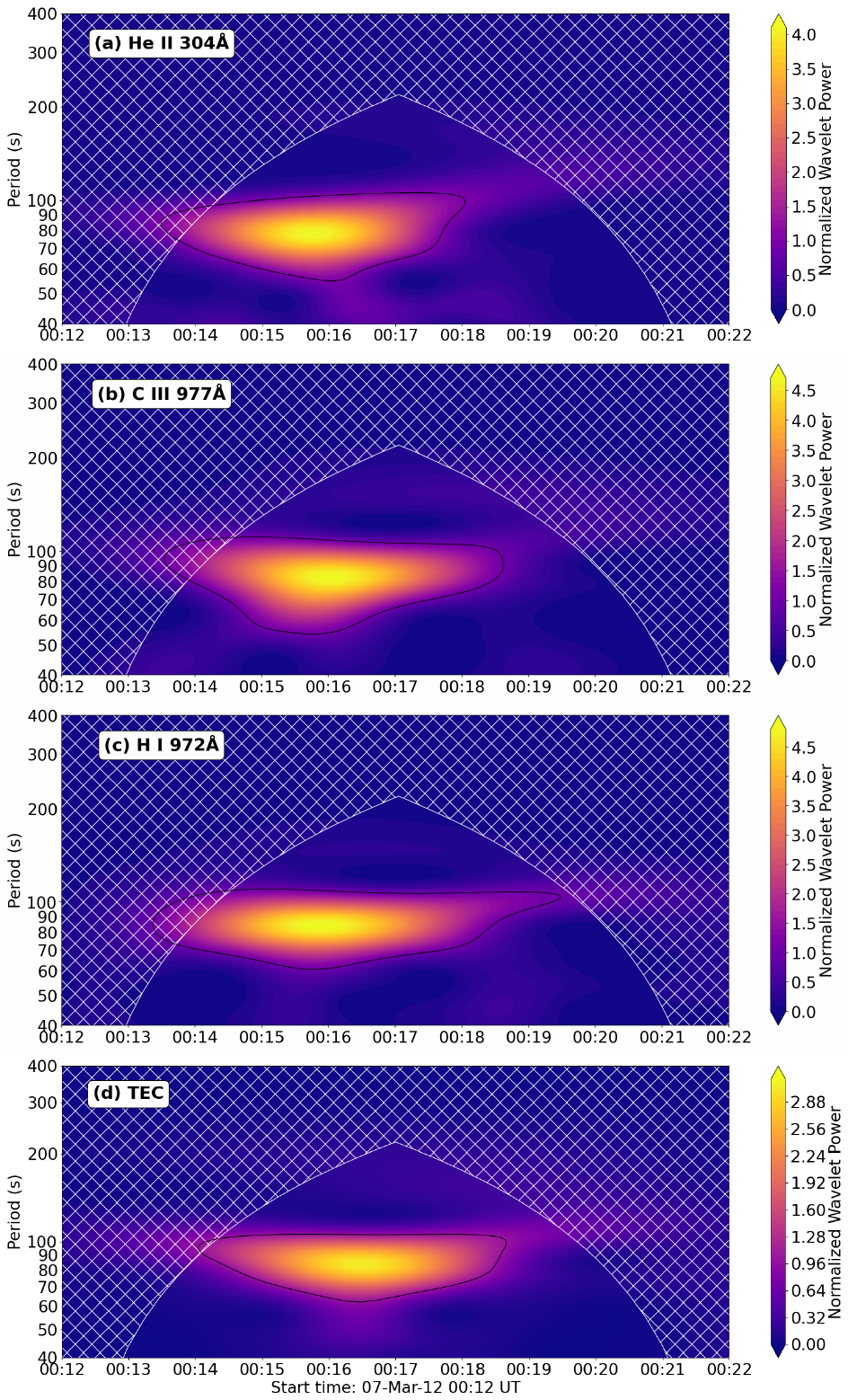}
    \caption{The wavelet power spectrum of the three EUV emission lines: (a) He~{\sc{ii}} 304\,\AA{}, (b) C~{\sc{iii}} 977\,\AA{}, (c) H~{\sc{i}} 972\,\AA{}, and (d) TEC. The solid black line denotes 99$\%$ significance. The white hashed area is outside the cone of influence, and the colour bar represents the normalized wavelet power.}
    \label{WPS}
\end{figure}

\begin{figure}[H]    
    \centering
    \includegraphics[width=.9\textwidth]{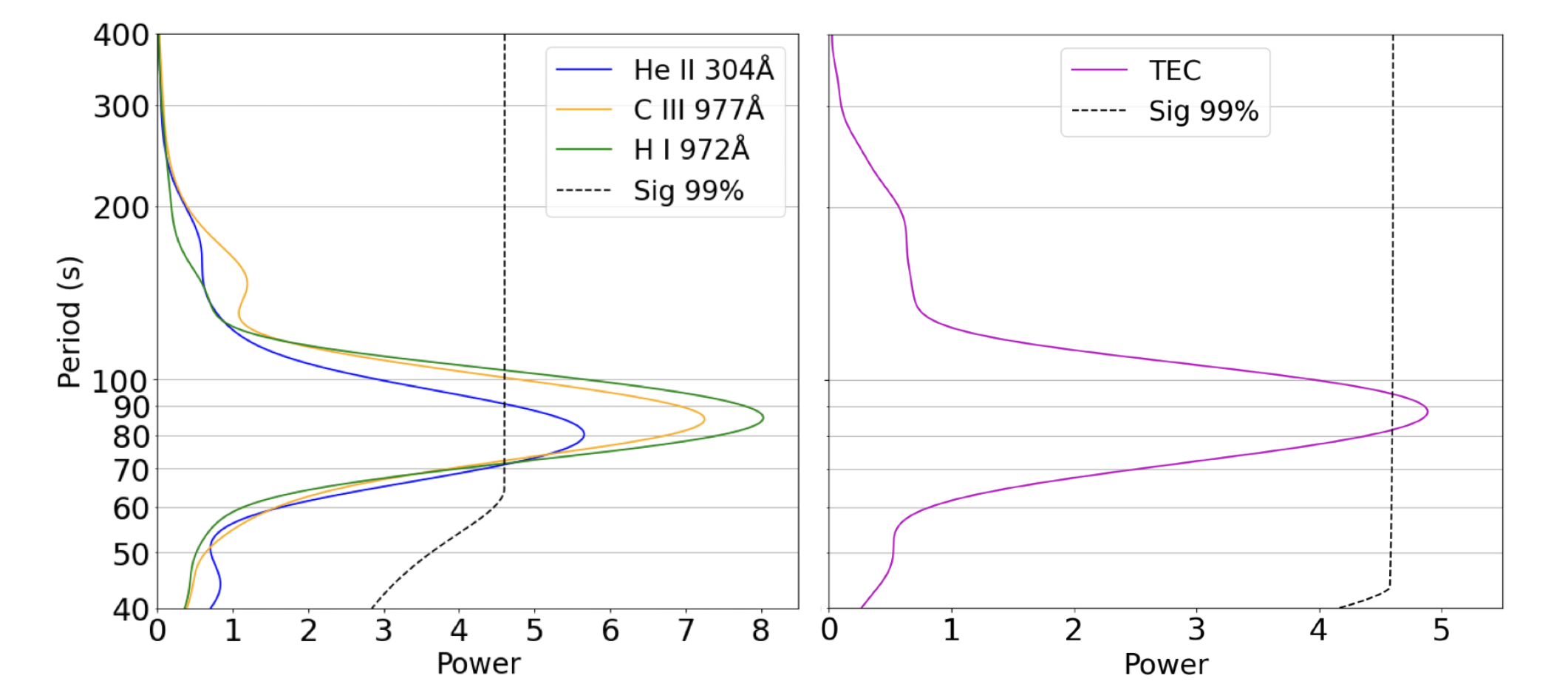}

    \caption{The global time-averaged wavelet power spectrum of the three EUV emission lines: He~{\sc{ii}} 304\,\AA{}, C~{\sc{iii}} 977\,\AA{}, H~{\sc{i}} 972\,\AA{} (left), and $\Delta$TEC (right). The dashed line in the global power spectrum is at the 99$\%$ significance level above the background model.}
    \label{GWS}
\end{figure}

Panels (a)–(c) of Figure \ref{WPS} show the results of the wavelet power spectrum for the the three EUV emission lines, He~{\sc{ii}} 304\,\AA{}, C~{\sc{iii}} 977\,\AA{} and H~{\sc{i}} 972\,\AA{}, respectively. Panel (d) of Figure \ref{WPS} shows the wavelet power spectrum for the $\Delta$TEC. Similarly, the left side of Figure \ref{GWS} shows the results of the global time-averaged wavelet power spectrum for the three EUV emission lines. The right side of Figure \ref{GWS} shows the global time-averaged wavelet power spectrum for the $\Delta$TEC. The black dashed line denotes the 99$\%$ significance level. As evident from Figures \ref{WPS} and \ref{GWS}, the significant timescales present in all three EUV lines found using wavelet analysis were consistent with those found for the $\Delta$TEC, and have an average of $\sim$85 seconds. 

\subsection{Periodogram Significance Testing}
\label{Periodogram}
\begin{figure}[h]    
    \centering
    \includegraphics[width=.9\textwidth]{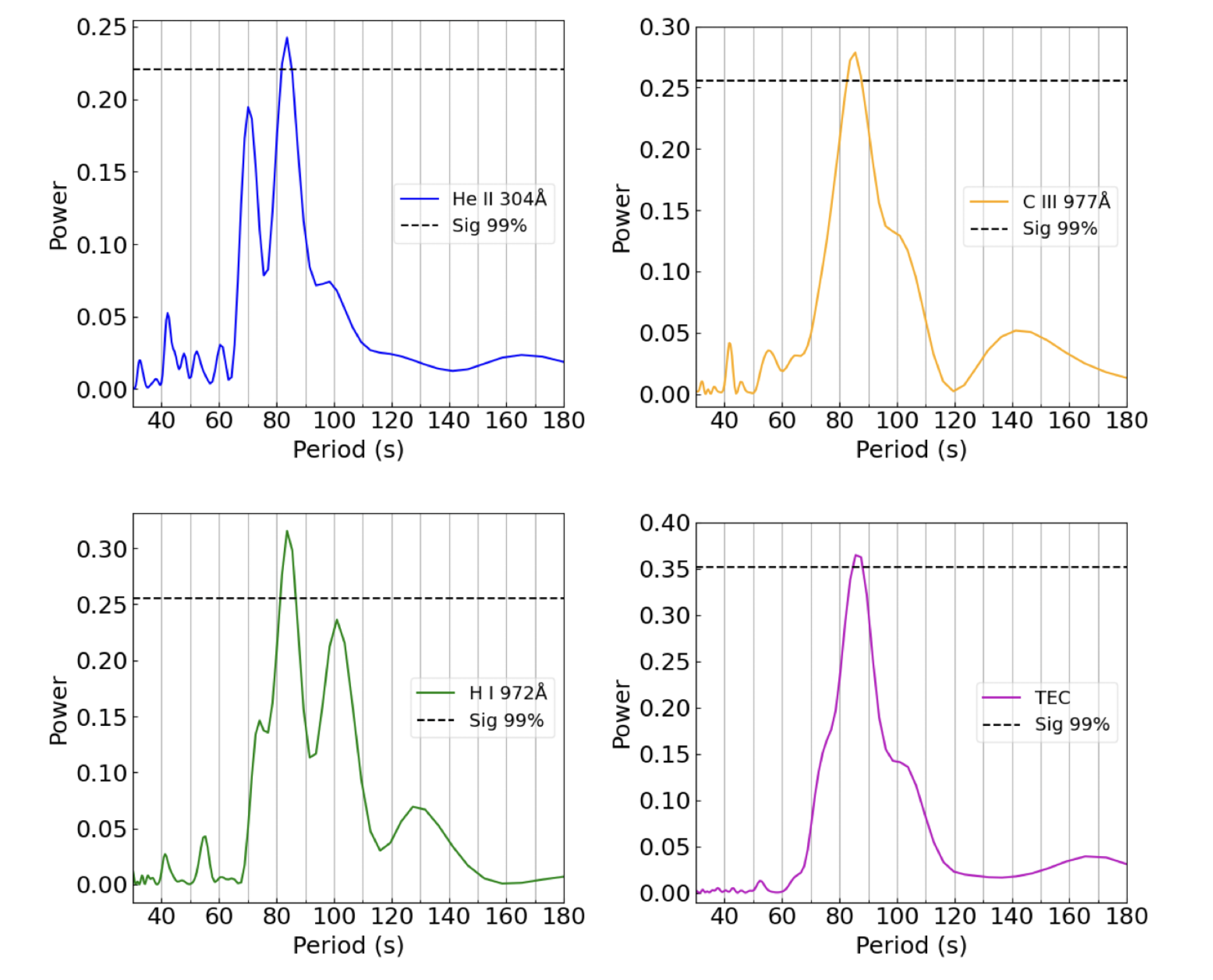}
    \caption{Lomb-Scargle periodograms for the three EUV emission lines: He~{\sc{ii}} 304\,\AA{} (blue), C~{\sc{iii}} 977\,\AA{} (orange), H~{\sc{i}} 972\,\AA{} (green), and TEC (magenta). The dashed lines denote the 99$\%$ significance level.}
    \label{periodograms}
\end{figure}

The Lomb–Scargle (LS) periodogram \citep{lomb1976least, scargle1982studies} is an algorithm for detecting periodicities in data by performing a Fourier-like transform to create a period–power spectrum. When using the LS periodogram to decide whether a signal contains a periodic component, an important consideration is the significance of the periodogram peak. This significance is expressed in terms of a false alarm probability, which encodes the probability of measuring a peak of a given height (or higher) conditioned on the assumption that the data consists of Gaussian noise with no periodic component. The false-alarm level, which is the required peak height to attain any given false alarm probability (e.g. false alarm level for a 1$\%$ false alarm probability is equivalent to a 99$\%$ significance level) was computed using the Astropy \citep{astropy:2013, astropy:2018, astropy:2022} Lomb-Scargle Periodogram functionality, and plotted on the periodogram to help identify the significant frequencies present. Figure \ref{periodograms} shows the LS periodograms for the three EUV emission lines: (a) He~{\sc{ii}} 304\,\AA{}, (b) C~{\sc{iii}} 977\,\AA{}, (c) H~{\sc{i}} 972\,\AA{}, and (d) $\Delta$TEC. The dashed line denotes the significance level of 99$\%$. The significant timescale ranges found using periodogram significance testing are narrower than those obtained using wavelet analysis, but they fully overlap and provide the same average period of $\sim$85 seconds. It is worth noting the pronounced peaks at $\sim$100\,s in the H~{\sc{i}} 972\,\AA{} periodogram and $\sim$70\,s in the He~{\sc{ii}} 304\,\AA{} periodogram. Although we see corresponding inflections in the $\Delta$TEC periodogram, since these peaks lie below the 99$\%$ significance level, they did not result in clear electron density pulsations in the ionosphere and therefore were not analyzed in this study.

\subsection{Cross-Correlation Analysis}
\label{crosscorr}
As previously mentioned, there is a time delay between the peaks in the detrended solar EUV emission lines and the detrended $\Delta$TEC curves. A value for this time delay was determined using cross-correlation analysis. A 30-second delay yielded the highest correlation coefficients ($r$) between the $\Delta$TEC and each EUV timeseries: $r$ = 0.78 for He~{\sc{ii}} 304\,\AA{}, $r$ = 0.88 for C~{\sc{iii}} 977\,\AA{}, and $r$ = 0.83 for H~{\sc{i}} 972\,\AA{}. This is expected given the delays visible in Figure \ref{detrTECEUV}. All three $r$ values calculated suggest a very strong association between the $\Delta$TEC and EUV timeseries.

\section{Results and Discussion}
We have detected and analyzed pulsations observed at multiple EUV wavelengths and in ionospheric TEC during the impulsive phase of the X5.4 solar flare on March 7, 2012. The range of characteristic timescales detected for the selected EUV lines and TEC found using both wavelet analysis and periodogram significance testing are listed in Table \ref{Table EUV}. 

Throughout the impulsive phase of this flare, highly correlated common features were observed in He~{\sc{ii}} 304\,\AA{}, C~{\sc{iii}} 977\,\AA{}, and H~{\sc{i}} 972\,\AA{}, with minimal time delay between peaks. Wavelet analysis of this impulsive interval (00:12–00:22\,UT) revealed broadband features in the wavelet power spectra across all three wavelengths of EUV emission, with similar power enhancements in each channel. The significant timescale ranges present in the time-averaged global power spectrum for the EUV fluxes were 71–90\,s (He~{\sc{ii}} 303.7\,\AA{}), 71–103\,s (H~{\sc{i}} 972\,\AA{}), and 72–100\,s (C~{\sc{iii}} 977\,\AA{}). The range of characteristic timescales found using periodogram analysis for the EUV fluxes were 82–85\,s (He~{\sc{ii}} 303.7\,\AA{}), 82–87\,s (H~{\sc{i}} 972\,\AA{}), and 83–88\,s (C~{\sc{iii}} 977\,\AA{}). These timescale ranges are consistent with those found using wavelet analysis.  

\begin{table}[!ht]
\caption{The significant timescale ranges of the oscillations found in the three geoeffective solar flare EUV emission lines (as measured by SDO/EVE), and TEC. The significant timescales are those for which the summed power exceeds the 99$\%$ significance level according to wavelet analysis and periodogram significance testing. }
\vspace{0.05\textwidth}
    \centering
    \begin{tabular}{lccc}
    \hline\hline
        Measurement & Significant Timescale & Significant Timescale \\
         & Range (s) [Wavelet] & Range (s) [Periodogram] \\
        \hline \hline
        Flux / He~{\sc{ii}} 303.7\,\AA{} & 71–90 & 82–85\\
        Flux / H~{\sc{i}}\;\;\;\;972.5\,\AA{} & 71–103 & 82–87\\
        Flux / C~{\sc{iii}} 977.0\,\AA{} & 72–100 & 83–88\\
        \hline
        TEC & 81–94 & 84–88 \\
        \hline
    \end{tabular}
    \label{Table EUV}
\end{table}

The averaged wavelet analysis of TEC timeseries data revealed very similar power enhancements in the wavelet power spectra as in the case of the three wavelengths of EUV emission. The average global spectrum for TEC revealed a significant timescale range between 81–94\,s. Periodogram significance testing of the TEC measurements revealed significant timescale ranges of pulsations between 84-88\,s. This is consistent with those found using wavelet analysis, as well as those found in all EUV emission lines. 

The time ranges found using periodogram significance testing are narrower than those returned by wavelet analysis. Additionally, the periodogram time ranges for the EUV lines are more similar to those found for TEC than the time ranges returned by wavelet analysis for the EUV lines. Based on this, periodogram significance testing is the preferred method for periodicity analysis in this study. However, wavelet analysis provides information about when in the interval the pulsations were present, and the use of more than one periodicity detection method is recommended for QPPs \cite{broomhall2019blueprint}. 
 
The high correlation coefficients ($r$) between the solar EUV emissions and $\Delta$TEC ($r$ = 0.78 for He~{\sc{ii}} 304\,\AA{}, $r$ = 0.88 for C~{\sc{iii}} 977\,\AA{}, and $r$ = 0.83 for H~{\sc{i}} 972\,\AA{}) indicate a strong relationship between the pulsations in the solar EUV emissions and the variations in ionospheric TEC. Peaks in ionospheric TEC were observed following those in the EUV flare emission, with a consistent time delay of approximately 30 seconds. This time delay is known as ionospheric ‘sluggishness' \cite{appleton1953note}, and is an inertial property of the ionosphere that is determined by the balance between ionization and recombination processes in the medium. Thus, the time delay depends on the dynamics of the ionization rate (the impulsiveness of the ionization source) and environmental parameters such as the solar zenith angle, the ionospheric altitude, latitude, background solar and magnetic activity. The magnitude of the obtained delay is consistent with previously reported values for various ionospheric layers \cite{hayes2020, Chakraborty2021, vzigman2023lower, bekker2024influence}, ranging from 45 seconds to several minutes. Notably, this is the first report of a time delay for TEC response to multiple peaks in individual emission lines from the same flare. Therefore, this approach can potentially be reliably used for the empirical estimation of the recombination rate in the F-region of the ionosphere.

\section{Conclusion}

This study presents the first recorded instance of synchronized pulsations in EUV flare emissions and ionospheric TEC, suggesting a highly sensitive coupling between oscillations in solar EUV radiation and the Earth's ionosphere on very short timescales. The selected EUV emission lines (He~{\sc{ii}} 304\,\AA{}, C~{\sc{iii}} 977\,\AA{}, and H~{\sc{i}} 972\,\AA{}) appear to significantly drive oscillatory behavior in TEC measurements, indicating a direct influence on ionospheric variability. However, further investigation is required to identify which specific emission lines modulate responses in individual ionospheric layers. As mentioned, \citet{Hayes2017} identified synchronized pulsations between solar flare X-ray emissions and VLF responses; however, these spanned multiple pulsations of GOES class B9.2–C6.8 and exhibited timescales on the order of tens of minutes. In contrast, the synchronized pulsations presented here occur on the shorter, more commonly observed timescales of tens of seconds, are present during the impulsive phase of a single flare, and have significantly smaller amplitudes than those in \citet{Hayes2017}. Furthermore, the time delay of 30\,s observed in this study between pulsations in solar EUV emissions and ionospheric TEC is the first reported for multiple synchronized, small-scale pulsations in individual solar EUV emission lines and TEC. This finding aligns with previously reported time delays between peaks in EUV flare lightcurves and the overall $\Delta$TEC peak during flare events \citep{bekker2024influence}. Additionally, the presented method for estimating the delay in TEC response to solar QPPs provides a practical tool for calculating the ionosphere's integral recombination rate. Given the short timescales of these pulsations, this approach allows for a more precise estimation of ionospheric sluggishness than methods based on longer-duration processes, such as gradual-onset flares.

The timescales of pulsations in EUV flare emission found in this study ($\sim$85\,s) are comparable with previously reported QPPs in solar flare events. For example, in a study of 90 $\geq$M5 class flares from Solar Cycle 24, \citet{dominique2018detection} found that 90$\%$ of the flares exhibited QPP periods between 1 and 100\,s in SXR and EUV wavelengths. Similarly, \citet{ning2017one} documented QPPs in SXRs with periods between 50 and 100\,s, while \citet{van2011lyra} identified a 63\,s period in Ly$\alpha$, a 74\,s period in the wavelength range 170\,\AA{}– 800\,\AA{} (including He~{\sc{ii}} 304\,\AA{} emission), and an 88\,s period for wavelengths between 60\,–200\,\AA{}. Pulsations within these timescale ranges could be the result of numerous driving mechanisms, including periodic reconnection, or modulation of electrons or emitting plasma by MHD waves \citep{Asai2001, Inglis2008, Ning2013, Nakariakov2016, Hayes2019}. Hard X-ray data were unavailable for this flare, thus it was not possible to conduct non-thermal electron and energy deposition diagnostics in order to determine the underlying driving mechanism of the solar flare pulsations. 

Future work should expand this analysis to a broader sample of flaring events, covering a range of intensities and spectral profiles, to assess how frequently this phenomenon occurs. An investigation should also be carried out to determine the driving mechanisms behind the pulsations in geoeffective flare emission and its location of origin on the Sun. Additionally, building on this work and the findings of \citeA{Hayes2017}, flares with concurrent X-ray and VLF (very low frequency) data should be investigated to identify pulsations occurring on timescales similar to those observed in this study, both in solar X-ray flux and in the resulting ionospheric D-region response. 

\section{Open Research}
The SDO/EVE data are publicly available at \url{https://lasp.colorado.edu/eve/data_access/index.html}. The GOES data can be accessed from \url{https://www.ngdc.noaa.gov/stp/satellite/goes-r.html}. Data from the SOPAC network are available at \url{http://sopac-old.ucsd.edu/.} The \citep{astropy:2013, astropy:2018, astropy:2022} Lomb-Scargle Periodogram documentation can be found at \url{https://docs.astropy.org/en/stable/api/astropy.timeseries.LombScargle.html#astropy.timeseries.LombScargle}. 

\section*{Acknowledgments}
A.O'H, S.Z.B. and R.O.M. would like to thank the European Office of Aerospace Research and Development (FA8655-22-1-7044P00001) for supporting this research. A.O'H would additionally like to thank the European Space Agency's Archival Research Visitor Programme for supporting this work. R.O.M. would also like to acknowledge support from the Science and Technology Facilities Council (STFC) grants ST/W001144/1 and ST/X000923/1. L.A.H was supported by an ESA Research Fellowship.

\bibliographystyle{agsm}
\bibliography{ref}

@article{wan2005gps,
  title={The GPS measured SITEC caused by the very intense solar flare on July 14, 2000},
  author={Wan, Weixing and Liu, Libo and Yuan, Hong and Ning, Baiqi and Zhang, Shunrong},
  journal={Advances in Space Research},
  volume={36},
  number={12},
  pages={2465--2469},
  year={2005},
  publisher={Elsevier}
}

@book{mitra1974ionospheric,
  title={Ionospheric effects of solar flares},
  author={Mitra, Ashesh Prasad},
  volume={46},
  year={1974},
  publisher={Springer}
}

@article{tsurutani2009brief,
  title={A brief review of “solar flare effects” on the ionosphere},
  author={Tsurutani, BT and Verkhoglyadova, OP and Mannucci, AJ and Lakhina, GS and Li, G and Zank, GP},
  journal={Radio Science},
  volume={44},
  number={01},
  pages={1--14},
  year={2009},
  publisher={AGU}
}

@article{davies1997studying,
  title={Studying the ionosphere with the Global Positioning System},
  author={Davies, K and Hartmann, GK},
  journal={Radio Science},
  volume={32},
  number={4},
  pages={1695--1703},
  year={1997},
  publisher={Wiley Online Library}
}

@inproceedings{leonovich2002estimating,
  title={Estimating the contribution from different ionospheric regions to the TEC response to the solar flares using data from the international GPS network},
  author={Leonovich, LA and Afraimovich, EL and Romanova, EB and Taschilin, AV},
  booktitle={Annales Geophysicae},
  volume={20},
  number={12},
  pages={1935--1941},
  year={2002},
  organization={Copernicus Publications G{\"o}ttingen, Germany}
}

@article{Nakariakov2009,
  doi = {10.1007/s11214-009-9536-3},
  year = {2009},
  month = may,
  publisher = {Springer Science and Business Media {LLC}},
  volume = {149},
  number = {1-4},
  pages = {119--151},
  author = {V. M. Nakariakov and V. F. Melnikov},
  title = {Quasi-Periodic Pulsations in Solar Flares},
  journal = {Space Science Reviews}
}

@article{simoes2015soft,
  title={Soft X-ray pulsations in solar flares},
  author={Sim{\~o}es, Paulo JA and Hudson, Hugh S and Fletcher, Lyndsay},
  journal={Solar Physics},
  volume={290},
  pages={3625--3639},
  year={2015},
  publisher={Springer}
}

@article{Hayes2017,
  doi = {10.1002/2017ja024647},
  year = {2017},
  publisher = {American Geophysical Union ({AGU})},
  volume = {122},
  number = {10},
  pages = {9841--9847},
  author = {Laura A. Hayes and Peter T. Gallagher and Joseph McCauley and Brian R. Dennis and Jack Ireland and Andrew Inglis},
  title = {Pulsations in the Earth{\textquotesingle}s Lower Ionosphere Synchronized With Solar Flare Emission},
  journal = {Journal of Geophysical Research: Space Physics}
}

@article{watanabe2021model,
  title={Model-based reproduction and validation of the total spectra of a solar flare and their impact on the global environment at the X9. 3 event of September 6, 2017},
  author={Watanabe, Kyoko and Jin, Hidekatsu and Nishimoto, Shohei and Imada, Shinsuke and Kawai, Toshiki and Kawate, Tomoko and Otsuka, Yuichi and Shinbori, Atsuki and Tsugawa, Takuya and Nishioka, Michi},
  journal={Earth, Planets and Space},
  volume={73},
  pages={1--10},
  year={2021},
  publisher={Springer}
}

@article{Solomon2005,
  title = {Solar extreme‐ultraviolet irradiance for general circulation models},
  volume = {110},
  ISSN = {0148-0227},
  url = {http://dx.doi.org/10.1029/2005JA011160},
  DOI = {10.1029/2005ja011160},
  number = {A10},
  journal = {Journal of Geophysical Research: Space Physics},
  publisher = {American Geophysical Union (AGU)},
  author = {Solomon,  Stanley C. and Qian,  Liying},
  year = {2005},
  month = oct 
}

@article{broomhall2019blueprint,
  title={A blueprint of state-of-the-art techniques for detecting quasi-periodic pulsations in solar and stellar flares},
  author={Broomhall, Anne-Marie and Davenport, James RA and Hayes, Laura A and Inglis, Andrew R and Kolotkov, Dmitrii Y and McLaughlin, James A and Mehta, Tishtrya and Nakariakov, Valery M and Notsu, Yuta and Pascoe, David J and others},
  journal={The Astrophysical Journal Supplement Series},
  volume={244},
  number={2},
  pages={44},
  year={2019},
  publisher={IOP Publishing}
}

@article{torrence1998practical,
  title={A practical guide to wavelet analysis},
  author={Torrence, Christopher and Compo, Gilbert P},
  journal={Bulletin of the American Meteorological society},
  volume={79},
  number={1},
  pages={61--78},
  year={1998},
  publisher={American Meteorological Society}
}

@article{lomb1976least,
  title={Least-squares frequency analysis of unequally spaced data},
  author={Lomb, Nicholas R},
  journal={Astrophysics and space science},
  volume={39},
  pages={447--462},
  year={1976},
  publisher={Springer}
}

@article{scargle1982studies,
  title={Studies in astronomical time series analysis. II-Statistical aspects of spectral analysis of unevenly spaced data},
  author={Scargle, Jeffrey D},
  journal={Astrophysical Journal, Part 1, vol. 263, Dec. 15, 1982, p. 835-853.},
  volume={263},
  pages={835--853},
  year={1982}
}

@article{dennis2017detection,
  title={Detection and interpretation of long-lived X-ray quasi-periodic pulsations in the X-class solar flare on 2013 May 14},
  author={Dennis, Brian R and Tolbert, Anne K and Inglis, Andrew and Ireland, Jack and Wang, Tongjiang and Holman, Gordon D and Hayes, Laura A and Gallagher, Peter T},
  journal={The Astrophysical Journal},
  volume={836},
  number={1},
  pages={84},
  year={2017},
  publisher={IOP Publishing}
}

@article{dolla2012time,
  title={Time delays in quasi-periodic pulsations observed during the X2. 2 solar flare on 2011 February 15},
  author={Dolla, Laurent and Marqu{\'e}, C and Seaton, DB and Van Doorsselaere, T and Dominique, M and Berghmans, D and Cabanas, C and De Groof, A and Schmutz, W and Verdini, Andrea and others},
  journal={The Astrophysical Journal Letters},
  volume={749},
  number={1},
  pages={L16},
  year={2012},
  publisher={IOP Publishing}
}

@article{dominique2018detection,
  title={Detection of quasi-periodic pulsations in solar EUV time series},
  author={Dominique, M and Zhukov, AN and Dolla, L and Inglis, A and Lapenta, G},
  journal={Solar Physics},
  volume={293},
  pages={1--24},
  year={2018},
  publisher={Springer}
}

@article{hayes2016quasi,
  title={Quasi-periodic pulsations during the impulsive and decay phases of an X-class flare},
  author={Hayes, Laura A and Gallagher, Peter T and Dennis, Brian R and Ireland, Jack and Inglis, Andrew R and Ryan, Daniel F},
  journal={The Astrophysical Journal Letters},
  volume={827},
  number={2},
  pages={L30},
  year={2016},
  publisher={IOP Publishing}
}

@book{pesnell2012solar,
  title={The solar dynamics observatory (SDO)},
  author={Pesnell, W Dean and Thompson, B Jꎬ and Chamberlin, PC},
  year={2012},
  publisher={Springer}
}

@article{savitzky1964smoothing,
  title={Smoothing and differentiation of data by simplified least squares procedures.},
  author={Savitzky, Abraham and Golay, Marcel JE},
  journal={Analytical chemistry},
  volume={36},
  number={8},
  pages={1627--1639},
  year={1964},
  publisher={ACS Publications}
}

@article{kupriyanova2010types,
  title={Types of microwave quasi-periodic pulsations in single flaring loops},
  author={Kupriyanova, EG and Melnikov, Victor Fedorovich and Nakariakov, VM and Shibasaki, Kiyoto},
  journal={Solar Physics},
  volume={267},
  pages={329--342},
  year={2010},
  publisher={Springer}
}

@article{parks1969sixteen,
  author="Parks, GK and Winckler, JR",
  title="Sixteen-second periodic pulsations observed in the correlated microwave and energetic X-ray emission from a solar flare",
  journal="Astrophysical Journal",
  volume="155",
  pages="L117",
  year="1969"
}

@article{li2015imaging,
  title={Imaging and spectral observations of quasi-periodic pulsations in a solar flare},
  author={Li, D and Ning, ZJ and Zhang, QM},
  journal={The Astrophysical Journal},
  volume={807},
  number={1},
  pages={72},
  year={2015},
  publisher={IOP Publishing}
}

@article{ning2017one,
  title={One-minute quasi-periodic pulsations seen in a solar flare},
  author={Ning, Z},
  journal={Solar Physics},
  volume={292},
  number={1},
  pages={11},
  year={2017},
  publisher={Springer}
}

@article{Milligan2017,
  doi = {10.3847/2041-8213/aa8f3a},
  year = {2017},
  month = oct,
  publisher = {American Astronomical Society},
  volume = {848},
  number = {1},
  pages = {L8},
  author = {Ryan O. Milligan and Bernhard Fleck and Jack Ireland and Lyndsay Fletcher and Brian R. Dennis},
  title = {Detection of Three-minute Oscillations in Full-disk Ly$\alpha$ Emission during a Solar Flare},
  journal = {The Astrophysical Journal}
}

@article{brosius2015quasi,
  title={Quasi-periodic fluctuations and chromospheric evaporation in a solar flare ribbon observed by IRIS},
  author={Brosius, Jeffrey W and Daw, Adrian N},
  journal={The Astrophysical Journal},
  volume={810},
  number={1},
  pages={45},
  year={2015},
  publisher={IOP Publishing}
}

@article{brosius2016quasi,
  title={Quasi-periodic fluctuations and chromospheric evaporation in a solar flare ribbon observed by Hinode/EIS, IRIS, and RHESSI},
  author={Brosius, Jeffrey W and Daw, Adrian N and Inglis, Andrew R},
  journal={The Astrophysical Journal},
  volume={830},
  number={2},
  pages={101},
  year={2016},
  publisher={IOP Publishing}
}

@article{tian2016global,
  title={Global sausage oscillation of solar flare loops detected by the interface region imaging spectrograph},
  author={Tian, Hui and Young, Peter R and Reeves, Katharine K and Wang, Tongjiang and Antolin, Patrick and Chen, Bin and He, Jiansen},
  journal={The Astrophysical Journal Letters},
  volume={823},
  number={1},
  pages={L16},
  year={2016},
  publisher={IOP Publishing}
}

@article{mclaughlin2018modelling,
  title={Modelling quasi-periodic pulsations in solar and stellar flares},
  author={McLaughlin, JA and Nakariakov, Valery M and Dominique, M and Jel{\'\i}nek, P and Takasao, S},
  journal={Space Science Reviews},
  volume={214},
  pages={1--54},
  year={2018},
  publisher={Springer}
}

@article{collier2024localising,
  title={Localising pulsations in the hard X-ray and microwave emission of an X-class flare},
  author={Collier, Hannah and Hayes, Laura A and Yu, Sijie and Battaglia, Andrea F and Ashfield, William and Polito, Vanessa and Harra, Louise K and Krucker, S{\"a}m},
  journal={Astronomy \& Astrophysics},
  volume={684},
  pages={A215},
  year={2024},
  publisher={EDP Sciences}
}

@article{thomson2001solar,
  title={Solar flare induced ionospheric D-region enhancements from VLF amplitude observations},
  author={Thomson, Neil R and Clilverd, Mark A},
  journal={Journal of Atmospheric and Solar-Terrestrial Physics},
  volume={63},
  number={16},
  pages={1729--1737},
  year={2001},
  publisher={Elsevier}
}

@article{raulin2013response,
  title={Response of the low ionosphere to X-ray and Lyman-$\alpha$ solar flare emissions},
  author={Raulin, Jean-Pierre and Trottet, G{\'e}rard and Kretzschmar, Matthieu and Macotela, Edith L and Pacini, Alessandra and Bertoni, Fernando CP and Dammasch, Ingolf E},
  journal={Journal of Geophysical Research: Space Physics},
  volume={118},
  number={1},
  pages={570--575},
  year={2013},
  publisher={Wiley Online Library}
}

@article{hayes2021solar,
  title={Solar flare effects on the earth’s lower ionosphere},
  author={Hayes, Laura A and O’Hara, Oscar SD and Murray, Sophie A and Gallagher, Peter T},
  journal={Solar Physics},
  volume={296},
  number={11},
  pages={157},
  year={2021},
  publisher={Springer}
}

@article{nina2021modelling,
  title={Modelling of the electron density and total electron content in the quiet and solar X-ray flare perturbed ionospheric d-region based on remote sensing by VLF/LF signals},
  author={Nina, Aleksandra},
  journal={Remote Sensing},
  volume={14},
  number={1},
  pages={54},
  year={2021},
  publisher={MDPI}
}

@article{bekker2023influence,
  title={Influence of the neutral atmosphere model on the correctness of simulation the electron and ion concentrations in the lower ionosphere},
  author={Bekker, SZ and Korsunskaya, JA},
  journal={Journal of Geophysical Research: Space Physics},
  volume={128},
  number={12},
  pages={e2023JA032007},
  year={2023},
  publisher={Wiley Online Library}
}

@article{wan2002sudden,
  title={The sudden increase in ionospheric total electron content caused by the very intense solar flare on July 14, 2000},
  author={Wan, Weixing and Yuan, Hong and Liu, Libo and Ning, Baiqi},
  journal={Science in China Series A: Mathematics},
  volume={45},
  pages={142--147},
  year={2002},
  publisher={Springer}
}

@article{garcia2007solar,
  title={Solar flare detection system based on global positioning system data: First results},
  author={Garcia-Rigo, A and Hern{\'a}ndez-Pajares, M and Juan, JM and Sanz, J},
  journal={Advances in Space Research},
  volume={39},
  number={5},
  pages={889--895},
  year={2007},
  publisher={Elsevier}
}

@article{yasyukevich20186,
  title={The 6 September 2017 X-class solar flares and their impacts on the ionosphere, GNSS, and HF radio wave propagation},
  author={Yasyukevich, Yu and Astafyeva, E and Padokhin, A and Ivanova, V and Syrovatskii, S and Podlesnyi, A},
  journal={Space Weather},
  volume={16},
  number={8},
  pages={1013--1027},
  year={2018},
  publisher={Wiley Online Library}
}

@article{knuth2020subsecond,
  title={Subsecond spikes in Fermi GBM X-ray flux as a probe for solar flare particle acceleration},
  author={Knuth, Trevor and Glesener, Lindsay},
  journal={The Astrophysical Journal},
  volume={903},
  number={1},
  pages={63},
  year={2020},
  publisher={IOP Publishing}
}

@article{van2011lyra,
  title={LYRA observations of two oscillation modes in a single flare},
  author={Van Doorsselaere, T and De Groof, A and Zender, J and Berghmans, D and Goossens, M},
  journal={The Astrophysical Journal},
  volume={740},
  number={2},
  pages={90},
  year={2011},
  publisher={IOP Publishing}
}

@article{Asai2001,
  title = {Periodic Acceleration of Electrons in the 1998 November 10 Solar Flare},
  volume = {562},
  ISSN = {0004-637X},
  url = {http://dx.doi.org/10.1086/338052},
  DOI = {10.1086/338052},
  number = {1},
  journal = {The Astrophysical Journal},
  publisher = {American Astronomical Society},
  author = {Asai,  A. and Shimojo,  M. and Isobe,  H. and Morimoto,  T. and Yokoyama,  T. and Shibasaki,  K. and Nakajima,  H.},
  year = {2001},
  month = nov,
  pages = {L103–L106}
}

@article{Inglis2008,
  title = {Multi-wavelength spatially resolved analysis of quasi-periodic pulsations in a solar flare},
  volume = {487},
  ISSN = {1432-0746},
  url = {http://dx.doi.org/10.1051/0004-6361:20079323},
  DOI = {10.1051/0004-6361:20079323},
  number = {3},
  journal = {Astronomy \& Astrophysics},
  publisher = {EDP Sciences},
  author = {Inglis,  A. R. and Nakariakov,  V. M. and Melnikov,  V. F.},
  year = {2008},
  month = jun,
  pages = {1147–1153}
}

@article{Nakariakov2016,
  title = {Magnetohydrodynamic Oscillations in the Solar Corona and Earth’s Magnetosphere: Towards Consolidated Understanding},
  volume = {200},
  ISSN = {1572-9672},
  url = {http://dx.doi.org/10.1007/s11214-015-0233-0},
  DOI = {10.1007/s11214-015-0233-0},
  number = {1–4},
  journal = {Space Science Reviews},
  publisher = {Springer Science and Business Media LLC},
  author = {Nakariakov,  V. M. and Pilipenko,  V. and Heilig,  B. and Jelínek,  P. and Karlický,  M. and Klimushkin,  D. Y. and Kolotkov,  D. Y. and Lee,  D.-H. and Nisticò,  G. and Van Doorsselaere,  T. and Verth,  G. and Zimovets,  I. V.},
  year = {2016},
  month = feb,
  pages = {75–203}
}

@article{Ning2013,
  title = {Imaging Observations of X-Ray Quasi-periodic Oscillations at 3–6 keV in the 26 December 2002 Solar Flare},
  volume = {289},
  ISSN = {1573-093X},
  url = {http://dx.doi.org/10.1007/s11207-013-0405-6},
  DOI = {10.1007/s11207-013-0405-6},
  number = {4},
  journal = {Solar Physics},
  publisher = {Springer Science and Business Media LLC},
  author = {Ning,  Zongjun},
  year = {2013},
  month = oct,
  pages = {1239–1256}
}

@article{Hayes2019,
  title = {Persistent Quasi-periodic Pulsations during a Large X-class Solar Flare},
  volume = {875},
  ISSN = {1538-4357},
  url = {http://dx.doi.org/10.3847/1538-4357/ab0ca3},
  DOI = {10.3847/1538-4357/ab0ca3},
  number = {1},
  journal = {The Astrophysical Journal},
  publisher = {American Astronomical Society},
  author = {Hayes,  Laura A. and Gallagher,  Peter T. and Dennis,  Brian R. and Ireland,  Jack and Inglis,  Andrew and Morosan,  Diana E.},
  year = {2019},
  month = apr,
  pages = {33}
}

@article{woods2012extreme,
  title={Extreme Ultraviolet Variability Experiment (EVE) on the Solar Dynamics Observatory (SDO): Overview of science objectives, instrument design, data products, and model developments},
  author={Woods, Thomas N and Eparvier, FG and Hock, R and Jones, AR and Woodraska, D and Judge, D and Didkovsky, L and Lean, J and Mariska, J and Warren, H and others},
  journal={The solar dynamics observatory},
  pages={115--143},
  year={2012},
  publisher={Springer}
}

@ARTICLE{hayes2020,
       author = {{Hayes}, Laura A. and {Inglis}, Andrew R. and {Christe}, Steven and {Dennis}, Brian and {Gallagher}, Peter T.},
        title = "{Statistical Study of GOES X-Ray Quasi-periodic Pulsations in Solar Flares}",
      journal = {The Astrophysical Journal},
         year = 2020,
        month = may,
       volume = {895},
       number = {1},
          eid = {50},
        pages = {50},
          doi = {10.3847/1538-4357/ab8d40},
archivePrefix = {arXiv},
       eprint = {2004.11775},
 primaryClass = {astro-ph.SR},
       adsurl = {https://ui.adsabs.harvard.edu/abs/2020ApJ...895...50H}
}

@article{Inglis2024,
  title = {Searching for Rapid Pulsations in Solar Flare X-Ray Data},
  volume = {971},
  ISSN = {1538-4357},
  url = {http://dx.doi.org/10.3847/1538-4357/ad54bb},
  DOI = {10.3847/1538-4357/ad54bb},
  number = {1},
  journal = {The Astrophysical Journal},
  publisher = {American Astronomical Society},
  author = {Inglis,  Andrew R. and Hayes,  Laura A.},
  year = {2024},
  month = aug,
  pages = {29}
}

@article{Zimovets2021,
  title = {Quasi-Periodic Energy Release in a Three-Ribbon Solar Flare},
  volume = {296},
  ISSN = {1573-093X},
  url = {http://dx.doi.org/10.1007/s11207-021-01936-9},
  DOI = {10.1007/s11207-021-01936-9},
  number = {12},
  journal = {Solar Physics},
  publisher = {Springer Science and Business Media LLC},
  author = {Zimovets,  Ivan and Sharykin,  Ivan and Myshyakov,  Ivan},
  year = {2021},
  month = dec 
}

@article{VanDoorsselaere2016,
  title = {Quasi-periodic Pulsations in Solar and Stellar Flares: An Overview of Recent Results (Invited Review)},
  volume = {291},
  ISSN = {1573-093X},
  url = {http://dx.doi.org/10.1007/s11207-016-0977-z},
  DOI = {10.1007/s11207-016-0977-z},
  number = {11},
  journal = {Solar Physics},
  publisher = {Springer Science and Business Media LLC},
  author = {Van Doorsselaere,  Tom and Kupriyanova,  Elena G. and Yuan,  Ding},
  year = {2016},
  month = sep,
  pages = {3143–3164}
}

@book{HofmannWellenhof1992,
  title = {Global Positioning System},
  ISBN = {9783709151266},
  url = {http://dx.doi.org/10.1007/978-3-7091-5126-6},
  DOI = {10.1007/978-3-7091-5126-6},
  publisher = {Springer Vienna},
  author = {Hofmann-Wellenhof,  Bernhard and Lichtenegger,  Herbert and Collins,  James},
  year = {1992}
}

@article{astropy:2013,
Adsurl = {http://adsabs.harvard.edu/abs/2013A%26A...558A..33A},
Archiveprefix = {arXiv},
Author = {{Astropy Collaboration} and {Robitaille}, T.~P. and {Tollerud}, E.~J. and {Greenfield}, P. and {Droettboom}, M. and {Bray}, E. and {Aldcroft}, T. and {Davis}, M. and {Ginsburg}, A. and {Price-Whelan}, A.~M. and {Kerzendorf}, W.~E. and {Conley}, A. and {Crighton}, N. and {Barbary}, K. and {Muna}, D. and {Ferguson}, H. and {Grollier}, F. and {Parikh}, M.~M. and {Nair}, P.~H. and {Unther}, H.~M. and {Deil}, C. and {Woillez}, J. and {Conseil}, S. and {Kramer}, R. and {Turner}, J.~E.~H. and {Singer}, L. and {Fox}, R. and {Weaver}, B.~A. and {Zabalza}, V. and {Edwards}, Z.~I. and {Azalee Bostroem}, K. and {Burke}, D.~J. and {Casey}, A.~R. and {Crawford}, S.~M. and {Dencheva}, N. and {Ely}, J. and {Jenness}, T. and {Labrie}, K. and {Lim}, P.~L. and {Pierfederici}, F. and {Pontzen}, A. and {Ptak}, A. and {Refsdal}, B. and {Servillat}, M. and {Streicher}, O.},
Doi = {10.1051/0004-6361/201322068},
Eid = {A33},
Eprint = {1307.6212},
Journal = {Astronomy \& Astrophysics},
Month = oct,
Pages = {A33},
Primaryclass = {astro-ph.IM},
Title = {{Astropy: A community Python package for astronomy}},
Volume = 558,
Year = 2013}

@ARTICLE{astropy:2018,
       author = {{Astropy Collaboration} and {Price-Whelan}, A.~M. and
         {Sip{\H{o}}cz}, B.~M. and {G{\"u}nther}, H.~M. and {Lim}, P.~L. and
         {Crawford}, S.~M. and {Conseil}, S. and {Shupe}, D.~L. and
         {Craig}, M.~W. and {Dencheva}, N. and {Ginsburg}, A. and {Vand
        erPlas}, J.~T. and {Bradley}, L.~D. and {P{\'e}rez-Su{\'a}rez}, D. and
         {de Val-Borro}, M. and {Aldcroft}, T.~L. and {Cruz}, K.~L. and
         {Robitaille}, T.~P. and {Tollerud}, E.~J. and {Ardelean}, C. and
         {Babej}, T. and {Bach}, Y.~P. and {Bachetti}, M. and {Bakanov}, A.~V. and
         {Bamford}, S.~P. and {Barentsen}, G. and {Barmby}, P. and
         {Baumbach}, A. and {Berry}, K.~L. and {Biscani}, F. and {Boquien}, M. and
         {Bostroem}, K.~A. and {Bouma}, L.~G. and {Brammer}, G.~B. and
         {Bray}, E.~M. and {Breytenbach}, H. and {Buddelmeijer}, H. and
         {Burke}, D.~J. and {Calderone}, G. and {Cano Rodr{\'\i}guez}, J.~L. and
         {Cara}, M. and {Cardoso}, J.~V.~M. and {Cheedella}, S. and {Copin}, Y. and
         {Corrales}, L. and {Crichton}, D. and {D'Avella}, D. and {Deil}, C. and
         {Depagne}, {\'E}. and {Dietrich}, J.~P. and {Donath}, A. and
         {Droettboom}, M. and {Earl}, N. and {Erben}, T. and {Fabbro}, S. and
         {Ferreira}, L.~A. and {Finethy}, T. and {Fox}, R.~T. and
         {Garrison}, L.~H. and {Gibbons}, S.~L.~J. and {Goldstein}, D.~A. and
         {Gommers}, R. and {Greco}, J.~P. and {Greenfield}, P. and
         {Groener}, A.~M. and {Grollier}, F. and {Hagen}, A. and {Hirst}, P. and
         {Homeier}, D. and {Horton}, A.~J. and {Hosseinzadeh}, G. and {Hu}, L. and
         {Hunkeler}, J.~S. and {Ivezi{\'c}}, {\v{Z}}. and {Jain}, A. and
         {Jenness}, T. and {Kanarek}, G. and {Kendrew}, S. and {Kern}, N.~S. and
         {Kerzendorf}, W.~E. and {Khvalko}, A. and {King}, J. and {Kirkby}, D. and
         {Kulkarni}, A.~M. and {Kumar}, A. and {Lee}, A. and {Lenz}, D. and
         {Littlefair}, S.~P. and {Ma}, Z. and {Macleod}, D.~M. and
         {Mastropietro}, M. and {McCully}, C. and {Montagnac}, S. and
         {Morris}, B.~M. and {Mueller}, M. and {Mumford}, S.~J. and {Muna}, D. and
         {Murphy}, N.~A. and {Nelson}, S. and {Nguyen}, G.~H. and
         {Ninan}, J.~P. and {N{\"o}the}, M. and {Ogaz}, S. and {Oh}, S. and
         {Parejko}, J.~K. and {Parley}, N. and {Pascual}, S. and {Patil}, R. and
         {Patil}, A.~A. and {Plunkett}, A.~L. and {Prochaska}, J.~X. and
         {Rastogi}, T. and {Reddy Janga}, V. and {Sabater}, J. and
         {Sakurikar}, P. and {Seifert}, M. and {Sherbert}, L.~E. and
         {Sherwood-Taylor}, H. and {Shih}, A.~Y. and {Sick}, J. and
         {Silbiger}, M.~T. and {Singanamalla}, S. and {Singer}, L.~P. and
         {Sladen}, P.~H. and {Sooley}, K.~A. and {Sornarajah}, S. and
         {Streicher}, O. and {Teuben}, P. and {Thomas}, S.~W. and
         {Tremblay}, G.~R. and {Turner}, J.~E.~H. and {Terr{\'o}n}, V. and
         {van Kerkwijk}, M.~H. and {de la Vega}, A. and {Watkins}, L.~L. and
         {Weaver}, B.~A. and {Whitmore}, J.~B. and {Woillez}, J. and
         {Zabalza}, V. and {Astropy Contributors}},
        title = "{The Astropy Project: Building an Open-science Project and Status of the v2.0 Core Package}",
      journal = {The Astrophysical Journal},
         year = 2018,
        month = sep,
       volume = {156},
       number = {3},
          eid = {123},
        pages = {123},
          doi = {10.3847/1538-3881/aabc4f},
archivePrefix = {arXiv},
       eprint = {1801.02634},
 primaryClass = {astro-ph.IM},
       adsurl = {https://ui.adsabs.harvard.edu/abs/2018AJ....156..123A}
}

@ARTICLE{astropy:2022,
       author = {{Astropy Collaboration} and {Price-Whelan}, Adrian M. and {Lim}, Pey Lian and {Earl}, Nicholas and {Starkman}, Nathaniel and {Bradley}, Larry and {Shupe}, David L. and {Patil}, Aarya A. and {Corrales}, Lia and {Brasseur}, C.~E. and {N{"o}the}, Maximilian and {Donath}, Axel and {Tollerud}, Erik and {Morris}, Brett M. and {Ginsburg}, Adam and {Vaher}, Eero and {Weaver}, Benjamin A. and {Tocknell}, James and {Jamieson}, William and {van Kerkwijk}, Marten H. and {Robitaille}, Thomas P. and {Merry}, Bruce and {Bachetti}, Matteo and {G{"u}nther}, H. Moritz and {Aldcroft}, Thomas L. and {Alvarado-Montes}, Jaime A. and {Archibald}, Anne M. and {B{'o}di}, Attila and {Bapat}, Shreyas and {Barentsen}, Geert and {Baz{'a}n}, Juanjo and {Biswas}, Manish and {Boquien}, M{'e}d{'e}ric and {Burke}, D.~J. and {Cara}, Daria and {Cara}, Mihai and {Conroy}, Kyle E. and {Conseil}, Simon and {Craig}, Matthew W. and {Cross}, Robert M. and {Cruz}, Kelle L. and {D'Eugenio}, Francesco and {Dencheva}, Nadia and {Devillepoix}, Hadrien A.~R. and {Dietrich}, J{"o}rg P. and {Eigenbrot}, Arthur Davis and {Erben}, Thomas and {Ferreira}, Leonardo and {Foreman-Mackey}, Daniel and {Fox}, Ryan and {Freij}, Nabil and {Garg}, Suyog and {Geda}, Robel and {Glattly}, Lauren and {Gondhalekar}, Yash and {Gordon}, Karl D. and {Grant}, David and {Greenfield}, Perry and {Groener}, Austen M. and {Guest}, Steve and {Gurovich}, Sebastian and {Handberg}, Rasmus and {Hart}, Akeem and {Hatfield-Dodds}, Zac and {Homeier}, Derek and {Hosseinzadeh}, Griffin and {Jenness}, Tim and {Jones}, Craig K. and {Joseph}, Prajwel and {Kalmbach}, J. Bryce and {Karamehmetoglu}, Emir and {Ka{l}uszy{'n}ski}, Miko{l}aj and {Kelley}, Michael S.~P. and {Kern}, Nicholas and {Kerzendorf}, Wolfgang E. and {Koch}, Eric W. and {Kulumani}, Shankar and {Lee}, Antony and {Ly}, Chun and {Ma}, Zhiyuan and {MacBride}, Conor and {Maljaars}, Jakob M. and {Muna}, Demitri and {Murphy}, N.~A. and {Norman}, Henrik and {O'Steen}, Richard and {Oman}, Kyle A. and {Pacifici}, Camilla and {Pascual}, Sergio and {Pascual-Granado}, J. and {Patil}, Rohit R. and {Perren}, Gabriel I. and {Pickering}, Timothy E. and {Rastogi}, Tanuj and {Roulston}, Benjamin R. and {Ryan}, Daniel F. and {Rykoff}, Eli S. and {Sabater}, Jose and {Sakurikar}, Parikshit and {Salgado}, Jes{'u}s and {Sanghi}, Aniket and {Saunders}, Nicholas and {Savchenko}, Volodymyr and {Schwardt}, Ludwig and {Seifert-Eckert}, Michael and {Shih}, Albert Y. and {Jain}, Anany Shrey and {Shukla}, Gyanendra and {Sick}, Jonathan and {Simpson}, Chris and {Singanamalla}, Sudheesh and {Singer}, Leo P. and {Singhal}, Jaladh and {Sinha}, Manodeep and {Sip{H{o}}cz}, Brigitta M. and {Spitler}, Lee R. and {Stansby}, David and {Streicher}, Ole and {{{S}}umak}, Jani and {Swinbank}, John D. and {Taranu}, Dan S. and {Tewary}, Nikita and {Tremblay}, Grant R. and {Val-Borro}, Miguel de and {Van Kooten}, Samuel J. and {Vasovi{'c}}, Zlatan and {Verma}, Shresth and {de Miranda Cardoso}, Jos{'e} Vin{'i}cius and {Williams}, Peter K.~G. and {Wilson}, Tom J. and {Winkel}, Benjamin and {Wood-Vasey}, W.~M. and {Xue}, Rui and {Yoachim}, Peter and {Zhang}, Chen and {Zonca}, Andrea and {Astropy Project Contributors}},
        title = "{The Astropy Project: Sustaining and Growing a Community-oriented Open-source Project and the Latest Major Release (v5.0) of the Core Package}",
      journal = {The Astrophysical Journal},
         year = 2022,
        month = aug,
       volume = {935},
       number = {2},
          eid = {167},
        pages = {167},
          doi = {10.3847/1538-4357/ac7c74},
archivePrefix = {arXiv},
       eprint = {2206.14220},
 primaryClass = {astro-ph.IM},
       adsurl = {https://ui.adsabs.harvard.edu/abs/2022ApJ...935..167A}
}

@article{Chakraborty2021,
  title = {Ionospheric Sluggishness: A Characteristic Time‐Lag of the Ionospheric Response to Solar Flares},
  volume = {126},
  ISSN = {2169-9402},
  url = {http://dx.doi.org/10.1029/2020JA028813},
  DOI = {10.1029/2020ja028813},
  number = {4},
  journal = {Journal of Geophysical Research: Space Physics},
  publisher = {American Geophysical Union (AGU)},
  author = {Chakraborty,  S. and Ruohoniemi,  J. M. and Baker,  J. B. H. and Fiori,  R. A. D. and Bailey,  S. M. and Zawdie,  K. A.},
  year = {2021},
  month = apr 
}

@article{bekker2024influence,
  title={The influence of different phases of a solar flare on changes in the total electron content in the Earth’s ionosphere},
  author={Bekker, Susanna and Milligan, Ryan O and Ryakhovsky, Ilya A},
  journal={The Astrophysical Journal},
  volume={971},
  number={2},
  pages={188},
  year={2024},
  publisher={IOP Publishing}
}

@article{appleton1953note,
  title={A note on the “sluggishness” of the ionosphere},
  author={Appleton, Edward V},
  journal={Journal of Atmospheric and Terrestrial Physics},
  volume={3},
  number={5},
  pages={282--284},
  year={1953},
  publisher={Elsevier}
}

@article{inglis2016large,
  title={A large-scale search for evidence of quasi-periodic pulsations in solar flares},
  author={Inglis, Andrew R. and Ireland, J and Dennis, BR and Hayes, L and Gallagher, P},
  journal={The Astrophysical Journal},
  volume={833},
  number={2},
  pages={284},
  year={2016},
  publisher={IOP Publishing}
}

@article{pugh2017properties,
  title={Properties of quasi-periodic pulsations in solar flares from a single active region},
  author={Pugh, Chloe E and Nakariakov, Valery M and Broomhall, A-M and Bogomolov, AV and Myagkova, IN},
  journal={Astronomy \& Astrophysics},
  volume={608},
  pages={A101},
  year={2017},
  publisher={EDP Sciences}
}

@article{vzigman2023lower,
  title={Lower-ionosphere electron density and effective recombination coefficients from multi-instrument space observations and ground VLF measurements during solar flares},
  author={{\v{Z}}igman, Vida and Dominique, Marie and Grubor, Davorka and Rodger, Craig J and Clilverd, Mark A},
  journal={Journal of Atmospheric and Solar-Terrestrial Physics},
  volume={247},
  pages={106074},
  year={2023},
  publisher={Elsevier}
}
\end{document}